\documentclass[10pt,conference]{IEEEtran}
\usepackage{multirow}
\usepackage{graphicx}
\usepackage{amssymb}
\usepackage{amsmath}
\usepackage{float}
\usepackage{tabularx}
\usepackage{caption}
\usepackage{fixltx2e} 
\graphicspath{ {./images/} }

\usepackage[sorting=none]{biblatex}
\begin{document}

\title{Open Vocabulary Keyword Spotting through Transfer Learning from Speech Synthesis}





\author{
\textit{Kesavaraj V, Anil Kumar Vuppala}\\
Speech Processing Laboratory, LTRC, \\
International Institute of Information Technology Hyderabad, India\\
\texttt{kesavaraj.v@research.iiit.ac.in, anil.vuppala@iiit.ac.in}\\
}
\maketitle

\begin{abstract}
Identifying keywords in an open-vocabulary context is crucial for personalizing interactions with smart devices.
Previous approaches to open vocabulary keyword spotting depend on a shared embedding space created by audio and text encoders. However, these approaches suffer from heterogeneous modality representations (i.e., audio-text mismatch).
To address this issue, our proposed framework leverages knowledge acquired from a pre-trained text-to-speech (TTS) system. This knowledge transfer allows for the incorporation of awareness of audio projections into the text representations derived from the text encoder. 
The performance of the proposed approach is compared with various baseline methods across four different datasets.
The robustness of our proposed model is evaluated by assessing its performance across different word lengths and in an Out-of-Vocabulary (OOV) scenario. Additionally, the effectiveness of
transfer learning from the TTS system is investigated by analyzing its different intermediate representations. 
The experimental results indicate that, in the challenging LibriPhrase Hard dataset, the proposed approach outperformed the cross-modality correspondence detector (CMCD) method 
by a significant improvement of 8.22\% in area under the curve (AUC) and 12.56\% in equal error rate (EER).





\end{abstract}


\section{Introduction}
Keyword spotting (KWS) is a process of detecting specific keywords within a continuous audio stream, which is crucial for enabling voice-driven interactions on edge devices \cite{lopez2021deep, tang2018deep}.
Increasing demand for personalized voice assistants highlights the significance of user-defined keyword spotting, which is known as custom keyword detection or open vocabulary keyword spotting \cite{sacchi2019open,kgurugubelli2024comparative}.
In contrast to closed vocabulary keyword spotting \cite{sainath15b_interspeech}, 
where only predetermined keywords are recognized,
custom keyword spotting, deals with the difficulty of identifying random keywords that the model might not have been exposed to during training, introducing an additional level of complexity to the task.

In the literature, there exist different methods for custom keyword spotting. One such method is the query-by-example (QbyE)~\cite{huang2021query, lugosch2018donut} approach, which involves matching input queries with pre-enrolled examples. However, the effectiveness of the QbyE method relies heavily on the similarity between the recorded speech during enrollment and the subsequent evaluated speech recordings. Factors like diverse vocal characteristics among users and environments with background noise pose significant challenges that can impact the consistency of performance of the QbyE method. In addressing this concern, researchers have delved into text enrollment-based methods \cite{sacchi2019open, shin22_interspeech}. One such method is the automatic speech recognition-based approach \cite{liu2021neural}, which detects phonetic patterns in input streams and then compares them with enrolled keyword representation. However, the effectiveness of this method heavily depends on the accuracy of the acoustic model. In \cite{shin22_interspeech}, an attention-based cross-modal matching approach is proposed that is trained in an end-to-end manner with monotonic matching loss and keyword classification loss. In \cite{lee23d_interspeech}, a zero-shot KWS is proposed, coupled with phoneme-level detection loss. Also, \cite{nishu23_interspeech} introduced Dynamic Sequence Partitioning to optimally partition the audio sequence to match the length of the word-based text sequence. These recent end-to-end techniques \cite{lee23d_interspeech, nishu23_interspeech, shin22_interspeech}, primarily depend on evaluating speech and text representations in a common latent space, demonstrated promising results in the custom keyword spotting task.
Therefore, we considered the end-to-end approach for our investigation.

\begin{figure*}[!htbp]
  \centering
  \includegraphics[width=\textwidth]{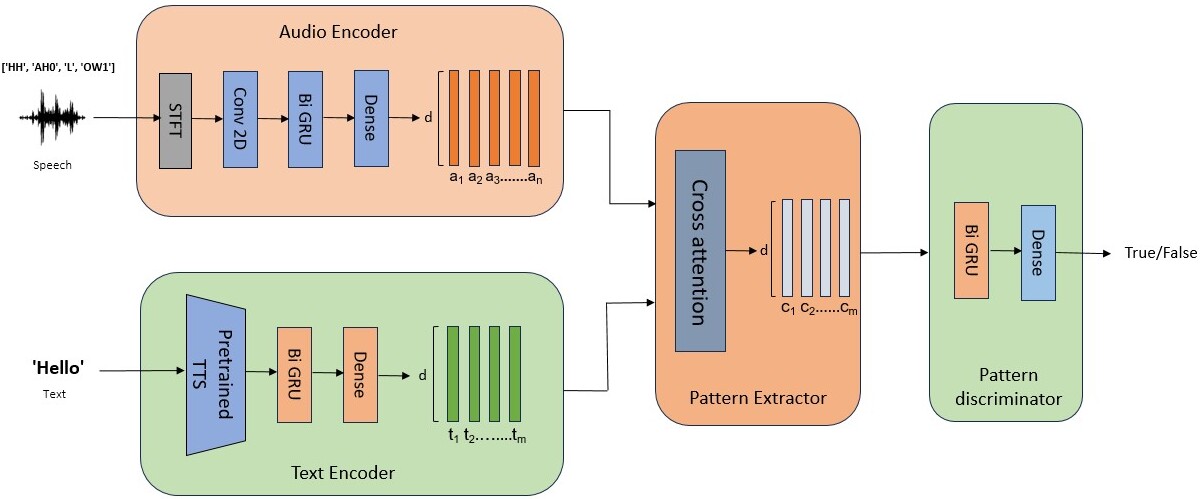}
  \caption{Proposed architecture for open vocabulary keyword spotting}
  \label{fig:architecture}
\end{figure*}
The effectiveness of the end-to-end approaches relies on the accuracy of projecting audio and text representations into a shared embedding space. While these methods \cite{shin22_interspeech, lee23d_interspeech} project speech and text representations into a shared latent space, they encounter difficulties in distinguishing between pairs of closely related pronunciations. This challenge can be addressed by generating text representations that have some acoustic knowledge. On the other hand, a TTS model converts text to audio, and the intermediate representations derived from this process possess an understanding of audio projections. Consequently, insights acquired from a pre-trained TTS model can serve as meaningful representations of text. In \cite{huang20i_interspeech}, a similar approach is used to transfer knowledge from a pre-trained TTS model for voice conversion task. Hence, in this paper, a novel framework is introduced to distill knowledge from a pre-trained TTS model for open-vocabulary keyword spotting task. The contributions of this study can be summarized as follows:
\begin{itemize}
  
  

\item A novel strategy is proposed that leverages intermediate representations extracted from a pre-trained TTS model as valuable text representations for custom KWS tasks.

\item Extensive investigation is conducted on various outputs of intermediate layers of the pre-trained TTS model to assess the efficacy of transfer learning. 


\item An ablation study is conducted to examine the system's performance for keywords of different word lengths.
\item Additionally, the robustness of the proposed framework is analyzed in an OOV scenario.

\end{itemize}

The following sections of the paper are organized as follows: Section~\ref{sec:Proposed method} introduces the proposed method in this study. Section~\ref{sec:Experimental Setup} provides details about the experimental setup, Section~\ref{sec:Results and Discussion} presents the results and discussion, and Section~\ref{sec:Conclusion} concludes the study.

\section{Proposed Method}
\label{sec:Proposed method}
This study introduces a novel methodology for open vocabulary keyword spotting by leveraging knowledge from a pre-trained TTS model. In this section, we describe the proposed architecture, as shown in Fig.~\ref{fig:architecture}. The architecture consists of four submodules: text encoder, audio encoder, pattern extractor, and pattern discriminator. 
The following sections contain detailed information about individual modules.


\vspace{-2pt}
\subsection{Text Encoder}


It consists of a pre-trained Tacotron 2 \cite{shen2018natural} model, a recurrent sequence-to-sequence TTS system, which takes character sequence as input for the corresponding keyword.
The resulting intermediate representations from the TTS model are then directed to a bidirectional gated recurrent unit (Bi-GRU) layer with a dimension of 64.
The dimensions of the intermediate representations from the TTS model depend on the specific layer from which they are extracted.
Further details of the intermediate representations are discussed in subsection~\ref{subsec:Tacotron 2}. 
The output from the Bi-GRU layer is then fed to a dense layer of size 128 units. The output from the text encoder is denoted as \(T_E \in \mathbb{R}^{d \times m}\), where d and m denote the embedding dimension and length of the text (i.e. number of characters), respectively.
 The primary motivation for incorporating the pre-trained TTS system into the text encoder is to generate text representations that are aware of acoustic projections. This integration simplifies the task of projecting audio and text embeddings in shared latent space. 
 Although this strategy does not explicitly involve generating audio from text, but exploits knowledge transfer from intermediate representations of TTS.
\vspace{-2pt}
\subsection{Audio Encoder}
The input to the audio encoder is 80-dimensional mel-filterbank coefficients, extracted for every 10ms with a 25ms window length. To process this audio feature, the encoder consists of two 2-D convolution layers (Conv 2D) with a kernel size of 3. The first convolution layer consists of 32 filters, and the second convolution layer consists of 64 filters. Batch normalization is applied after each convolution layer to ensure stable training.  To improve computational efficiency, the initial convolutional layer employs a stride of 2 to effectively reduce the number of processed frames by skipping consecutive frames. Following the convolution layers, two Bi-GRU layers each having a dimension of 64 are employed. Subsequently, the output from the last Bi-GRU layer is passed into a dense layer which generates a 128-dimensional audio embedding. 
The output from the audio encoder is denoted as
\(A_E \in \mathbb{R}^{d \times n}\), where d and n denote the embedding dimension and length of the audio (i.e. number of frames), respectively.

\vspace{-2pt}


\renewcommand{\arraystretch}{1.5}



\begin{table*}[!t]
\caption{Performance comparison of the proposed method with various KWS techniques across different datasets: Google Commands V1 (G), Qualcomm Keyword Speech dataset (Q), LibriPhrase-Easy (LP\textsubscript{E}), and LibriPhrase-Hard (LP\textsubscript{H}).
}
\setlength{\tabcolsep}{10 pt}
\label{tab:baseline comparison}
\small
\centering
\begin{tabular}{|c|cccc|cccc|}

\hline
\multirow{2}{*}{\textbf{Method}} & \multicolumn{4}{c|}{\textbf{EER (\%)}}                                                                                         & \multicolumn{4}{c|}{\textbf{AUC (\%)}}                                                                                           \\ \cline{2-9} 
                                 & \multicolumn{1}{c|}{\textbf{G}}    & \multicolumn{1}{c|}{\textbf{Q}}     & \multicolumn{1}{c|}{\textbf{LP\textsubscript{E}}}  & \textbf{LP\textsubscript{H}}   & \multicolumn{1}{c|}{\textbf{G}}     & \multicolumn{1}{c|}{\textbf{Q}}     & \multicolumn{1}{c|}{\textbf{LP\textsubscript{E}}}   & \textbf{LP\textsubscript{H}}   \\ \hline
CTC \cite{lugosch2018donut}                             & \multicolumn{1}{c|}{31.65}         & \multicolumn{1}{c|}{18.23}          & \multicolumn{1}{c|}{14.67}         & 35.22          & \multicolumn{1}{c|}{66.36}          & \multicolumn{1}{c|}{89.69}          & \multicolumn{1}{c|}{92.29}          & 69.58          \\ \hline
Attention \cite{huang2021query}                       & \multicolumn{1}{c|}{\textbf{14.75}}         & \multicolumn{1}{c|}{49.13}          & \multicolumn{1}{c|}{28.74}         & 41.95          & \multicolumn{1}{c|}{\textbf{92.09}}          & \multicolumn{1}{c|}{50.13}          & \multicolumn{1}{c|}{78.74}          & 62.65          \\ \hline
Triplet \cite{sacchi2019open}                          & \multicolumn{1}{c|}{35.6}          & \multicolumn{1}{c|}{38.72}          & \multicolumn{1}{c|}{32.75}         & 44.36          & \multicolumn{1}{c|}{71.48}          & \multicolumn{1}{c|}{66.44}          & \multicolumn{1}{c|}{63.53}          & 54.88          \\ \hline
CMCD \cite{shin22_interspeech}                            & \multicolumn{1}{c|}{27.25}         & \multicolumn{1}{c|}{12.15}          & \multicolumn{1}{c|}{8.42}          & 32.9           & \multicolumn{1}{c|}{81.06}          & \multicolumn{1}{c|}{94.51}          & \multicolumn{1}{c|}{96.7}           & 73.58          \\ \hline
\textbf{Proposed}                & \multicolumn{1}{c|}{22.3} & \multicolumn{1}{c|}{\textbf{10.82}} & \multicolumn{1}{c|}{\textbf{5.61}} & \textbf{24.68} & \multicolumn{1}{c|}{85.16} & \multicolumn{1}{c|}{\textbf{95.65}} & \multicolumn{1}{c|}{\textbf{98.49}} & \textbf{86.14} \\ \hline
\end{tabular}%
\vspace{-0.25cm}
\end{table*}


\subsection{Pattern Extractor}
It is based on the cross-attention mechanism \cite{vaswani2017attention} which captures the temporal correlations between the audio and text embeddings. 
In this setup, the audio embedding $A_E$ functions as both the key and value, while the text embedding $T_E$ acts as the query. The output of the pattern extractor is the context vector which contains information about audio and text agreement.


\subsection{Pattern Discriminator}
It consists of a single Bi-GRU layer with a dimension of 128 that takes the context vector from the cross-attention layer as input. Output from the last frame of the Bi-GRU layer is fed into a dense layer with a sigmoid as an activation function. The pattern discriminator determines whether audio and text inputs share the same keyword or not.

\section{Experimental Setup}
This section describes the database, Tacotron 2 embeddings, and implementation details for training.
\label{sec:Experimental Setup}
\subsection{Database}

For training and evaluation, we used the Libriphrase \cite{shin22_interspeech} dataset, which comprises short phrases with varying word lengths (1 to 4). 
The dataset was generated from the Librispeech corpus \cite{panayotov2015librispeech}.
The training set of Libriphrase is generated using train-clean-100 and train-clean-360 subsets, and the evaluation set is using train-others-500 subset. The evaluation set consists of 4391, 2605, 467, and 56 episodes of each word length respectively. Each episode has three positive and three negative pairs. The negative samples are further categorized into easy and hard based on Levenshtein distance \cite{levenshtein1966binary}, leading to the creation of the Libriphrase Easy (LP\textsubscript{E}) and Libriphrase Hard (LP\textsubscript{H}) datasets.  Each sample in the dataset is represented by using three entities: audio, text, and a binary target value indicating 1 for a positive pair and 0 for a negative pair.


To comprehensively evaluate model performance, we extended our assessment beyond the Libriphrase dataset by incorporating the Google Speech Commands V1 dataset (G) \cite{warden2018speech} and the Qualcomm Keyword Speech dataset (Q) \cite{kim2019query}. The Google Speech Commands V1 dataset (G) encompasses speech recordings from 1,881 speakers, focusing on 30 small keywords. 
On the other hand, the Qualcomm Keyword Speech dataset (Q) comprises 4,270 utterances of four keywords spoken by 50 speakers. 

\subsection{Tacotron 2 Embeddings}
\label{subsec:Tacotron 2}
In this study, the utilization of intermediate representation from Tacotron 2 is proposed to enhance performance, particularly in scenarios involving speech-keyword pairs with similar pronunciations. The intermediate representations are obtained using NVIDIA Tacotron 2 model \cite{NVIDIATacotron2} which is pre-trained on the LJSpeech dataset \cite{ito2017ljspeech}. The details of different intermediate representations that are used in this study are outlined in Table~\ref{tab:intermediate-layers}. Additional insights into the Tacotron 2 architecture can be found in~\cite{shen2018natural}





\begin{table}[ht]
\centering
\caption{Description of intermediate representations of Tacotron 2 model. $T_a-\text{number of characters}$, $T_b-\text{number of frames derived from the Tacotron 2}$.}
\label{tab:intermediate-layers}
\small
\begin{tabular}{|c|c|c|}
\hline
\textbf{Embedding} & \textbf{ Dimension} & \textbf{Description}       \\ \hline
E1                            & $T_a \times 512$                     & CharEmbedding block output \\ \hline
E2                            & $T_a \times 512$                        & Convolution block output   \\ \hline
E3                            & $T_a \times 512$                       & Bi-LSTM block output       \\ \hline
E4                            & $T_b \times 512$                      & Attention block output     \\ \hline
E5                            & $T_b \times 512$                       & Prenet block output        \\ \hline
E6                            & $T_b \times 512$                        & Postnet block output       \\ \hline
E7                            & $T_b \times 80$                         & Target Melspectrogram      \\ \hline
\end{tabular}%
\vspace{-0.25cm}
\end{table}




\begin{table*}[!ht]
\caption{Evaluation of different intermediate representations of Tacotron 2 model for custom KWS task across different datasets: Google Commands V1 dataset (G), Qualcomm Keyword Speech dataset (Q), LibriPhrase-Easy dataset (LP\textsubscript{E}), and LibriPhrase-Hard dataset (LP\textsubscript{H})}
\label{tab:ablation study}
\small
\centering
\begin{tabular}{|c|cccc|cccc|cccc|}
\hline
\multirow{2}{*}{\textbf{Method}} & \multicolumn{4}{c|}{\textbf{EER (\%)}}                                                                                         & \multicolumn{4}{c|}{\textbf{AUC (\%)}}                                                                                           & \multicolumn{4}{c|}{\textbf{F1 score (\%)}}                                                                                    \\ \cline{2-13} 
                                 & \multicolumn{1}{c|}{\textbf{G}}    & \multicolumn{1}{c|}{\textbf{Q}}     & \multicolumn{1}{c|}{\textbf{LP\textsubscript{E}}}  & \textbf{LP\textsubscript{H}}   & \multicolumn{1}{c|}{\textbf{G}}     & \multicolumn{1}{c|}{\textbf{Q}}     & \multicolumn{1}{c|}{\textbf{LP\textsubscript{E}}}   & \textbf{LP\textsubscript{H}}   & \multicolumn{1}{c|}{\textbf{G}}     & \multicolumn{1}{c|}{\textbf{Q}}     & \multicolumn{1}{c|}{\textbf{LP\textsubscript{E}}}  & \textbf{LP\textsubscript{H}}  \\ \hline
E1            & \multicolumn{1}{c|}{31.44}         & \multicolumn{1}{c|}{19.32}          & \multicolumn{1}{c|}{9.26}          & 32.13          & \multicolumn{1}{c|}{74.84}          & \multicolumn{1}{c|}{88.77}          & \multicolumn{1}{c|}{96.29}          & 74.67          & \multicolumn{1}{c|}{69.23}          & \multicolumn{1}{c|}{81.45}          & \multicolumn{1}{c|}{91.2}          & 64.86         \\ \hline
E2          & \multicolumn{1}{c|}{26.43}         & \multicolumn{1}{c|}{14.52}          & \multicolumn{1}{c|}{7.11}          & 28.47          & \multicolumn{1}{c|}{80.66}          & \multicolumn{1}{c|}{93.02}          & \multicolumn{1}{c|}{97.67}          & 79.6           & \multicolumn{1}{c|}{74.2}           & \multicolumn{1}{c|}{79.83}          & \multicolumn{1}{c|}{93.26}         & 67.78         \\ \hline
E3                 & \multicolumn{1}{c|}{\textbf{22.3}} & \multicolumn{1}{c|}{\textbf{10.82}} & \multicolumn{1}{c|}{\textbf{5.61}} & \textbf{24.68} & \multicolumn{1}{c|}{\textbf{85.16}} & \multicolumn{1}{c|}{\textbf{95.65}} & \multicolumn{1}{c|}{\textbf{98.49}} & \textbf{86.14} & \multicolumn{1}{c|}{\textbf{78.56}} & \multicolumn{1}{c|}{\textbf{87.74}} & \multicolumn{1}{c|}{\textbf{94.5}} & \textbf{72.2} \\ \hline
E4            & \multicolumn{1}{c|}{24.6}          & \multicolumn{1}{c|}{15.27}          & \multicolumn{1}{c|}{8.43}          & 31.2           & \multicolumn{1}{c|}{81.83}          & \multicolumn{1}{c|}{90.63}          & \multicolumn{1}{c|}{96.61}          & 76.25          & \multicolumn{1}{c|}{74.77}          & \multicolumn{1}{c|}{68.23}          & \multicolumn{1}{c|}{91.7}          & 67.54         \\ \hline
E5               & \multicolumn{1}{c|}{27.51}         & \multicolumn{1}{c|}{19.98}          & \multicolumn{1}{c|}{12.02}         & 34.35          & \multicolumn{1}{c|}{79.71}          & \multicolumn{1}{c|}{87.55}          & \multicolumn{1}{c|}{94.5}           & 71.77          & \multicolumn{1}{c|}{66.42}          & \multicolumn{1}{c|}{72.58}          & \multicolumn{1}{c|}{86.94}         & 71.41         \\ \hline
E6              & \multicolumn{1}{c|}{34.61}         & \multicolumn{1}{c|}{33.14}          & \multicolumn{1}{c|}{20.9}          & 39.19          & \multicolumn{1}{c|}{69.95}          & \multicolumn{1}{c|}{72.81}          & \multicolumn{1}{c|}{87.06}          & 65.29          & \multicolumn{1}{c|}{66.32}          & \multicolumn{1}{c|}{66.85}          & \multicolumn{1}{c|}{78.51}         & 61.47         \\ \hline
E7       & \multicolumn{1}{c|}{25.42}         & \multicolumn{1}{c|}{23.79}          & \multicolumn{1}{c|}{11.68}         & 34.7           & \multicolumn{1}{c|}{81.14}          & \multicolumn{1}{c|}{84.5}           & \multicolumn{1}{c|}{94.57}          & 70.97          & \multicolumn{1}{c|}{73.14}          & \multicolumn{1}{c|}{71.2}           & \multicolumn{1}{c|}{87.5}          & 66.78         \\ \hline
\end{tabular}%
\end{table*}

\subsection{Implementation Details}

The training pipeline is structured as a binary classification task, wherein the model is tasked with classifying the similarity of input pairs \{text, audio\}.
In audio and text encoders, weights are initialized through Xavier initialization, and Leaky ReLU is utilized as the activation function. A dropout of 0.2 is added after each layer in the encoders to prevent overfitting.
The training process uses binary cross-entropy loss as the training criterion and employs the Adam optimizer \cite{kingma2014adam} with default parameters for optimization.
A fixed learning rate of $10^{-4}$ and a batch size of 128 is employed in the training process. The best-performing model was chosen based on its performance on the validation set. For training, we used four NVIDIA GeForce RTX 2080 Ti GPUs.



\section{Results and Discussion}
\label{sec:Results and Discussion}
The proposed method explores intermediate representations from a pre-trained TTS model to boost the initialization of the text encoder. By tapping into the audio projections within the pre-trained TTS model, we aim to enrich text representations for custom keyword spotting task. 
We conducted extensive experiments across diverse datasets to evaluate this approach, with results presented in Tables -~\ref{tab:baseline comparison},~\ref{tab:ablation study},~\ref{tab:word length},~\ref{tab:OOV}.

Table~\ref{tab:baseline comparison} presents the comparative analysis of baselines and proposed approach across G, Q, LP\textsubscript{E}, and LP\textsubscript{H} datasets. Evaluation results show that among all the baselines CMCD demonstrates strong performance and Triplet shows weak performance across the Q, LP\textsubscript{E}, and LP\textsubscript{H} datasets, in terms of AUC and EER.  On the other hand,
the attention-based QbyE method shows powerful performance on the G dataset, which consists of frequently used words (e.g., "on", "off") as keywords. This method benefits when the keyword is part of the training set due to its similarity scoring mechanism, but it shows degraded performance when the keyword is unfamiliar, as observed in Q and LibriPhrase.
 However, our proposed approach outperforms all the baselines on all datasets except G.
 It is also evident, in comparison with the CMCD baseline, the proposed method showcases a significant improvement of 8.22 \% in AUC and 12.56\% in EER on the challenging LP\textsubscript{H} dataset which consists of similar audio-text pairs (e.g., "madame" and "modem"). This substantial improvement highlights the effectiveness of our method in better discrimination of closely related pronunciations. Moreover, we measure the generalization of the model on G and Q datasets, without any finetuning.  We find a consistent improvement of around 3\% on the AUC metric and 2.62\% on the EER metric across G and Q datasets when compared with the CMCD baseline.



Further, to assess the efficacy of transfer learning from pre-trained TTS for custom KWS task, an ablation study is conducted. This study compares embeddings obtained from various intermediate layers of the Tacotron-2 model, and results of this comparison are reported in Table~\ref{tab:ablation study}. 
Analyzing the results, E3 consistently outperforms others in terms of lower Equal Error Rate (EER) and higher AUC and F1-score across all datasets. This suggests it captures both acoustic and linguistic information of the enrolled keyword more effectively. Moreover, E4 and E2 exhibit competitive performance with consistently good AUC scores, signifying them as better alternatives to E3. Conversely, E6 shows poor performance compared to other layer embeddings, indicating its limitations for this task. 
Additionally, it can be inferred that intermediate representations (E2, E3) from the Tacotron~2 encoder seem to be significantly more suitable for the custom KWS task in comparison to representations (E5, E6, E7) from the Tacontron~2 decoder. This implies that capturing information before mel-spectrogram generation in the encoder stage is crucial for accurate keyword detection.

\begin{table}[!htbp]
\small
\centering
\caption{Performance of the proposed approach across different word lengths}
\vspace{5pt}
\label{tab:word length}
\begin{tabular}{|c|c|c|c|}
\hline
\textbf{Word length} & \textbf{EER (\%)} & \textbf{AUC (\%)} & \textbf{F1 score (\%)} \\ \hline
1                    & 5.41              & 98.07             & 94.46                  \\ \hline
2                    & 5.9               & 97.83             & 94.24                     \\ \hline
3                    & 7.59              & 97.04             & 92.28                  \\ \hline
4                    & 8.5               & 97.12             & 90.85                  \\ \hline
\end{tabular}%
\vspace{-0.4cm}
\end{table}

Table~\ref{tab:word length} presents the performance analysis of the proposed system across various word lengths. The evaluation of a system across different word lengths (1 to 4) reveals that shorter words (1 or 2) result in better performance, with lower EER and higher F1-score suggesting higher accuracy in keyword identification. As word length increases, EER values rise, indicating increased difficulty in recognition. 
However, our method exhibits consistent performance across all word lengths.
Additionally, to assess the robustness of the proposed system, we compared its performance in the OOV scenario. From Table~\ref{tab:OOV}, it is evident that,  in comparison to the CMCD baseline method, the proposed method shows an absolute improvement of 7.25\%, 6.36\%, and 5.53\% in terms of F1 score, AUC and EER, respectively. 
This signifies the effectiveness of the proposed system in handling user-defined keywords that are not seen during training. 



\begin{table}[!htbp]
\caption{Comparison of CMCD and proposed approach in Out-of-Vocabulary scenario.}
\vspace{5pt}
\label{tab:OOV}
\small
\centering
\begin{tabular}{|c|c|c|c|}
\hline
\textbf{Method} & \textbf{EER (\%)} & \textbf{AUC (\%)} & \textbf{F1 score (\%)} \\ \hline
CMCD        & 23.48             & 84.08             & 76.2                   \\ \hline
Proposed        & \textbf{18.14}    & \textbf{90.44}    & \textbf{83.45}         \\ \hline
\end{tabular}%
\end{table}

\begin{figure}[!t]
  \centering
  \resizebox{\linewidth}{!}{\includegraphics{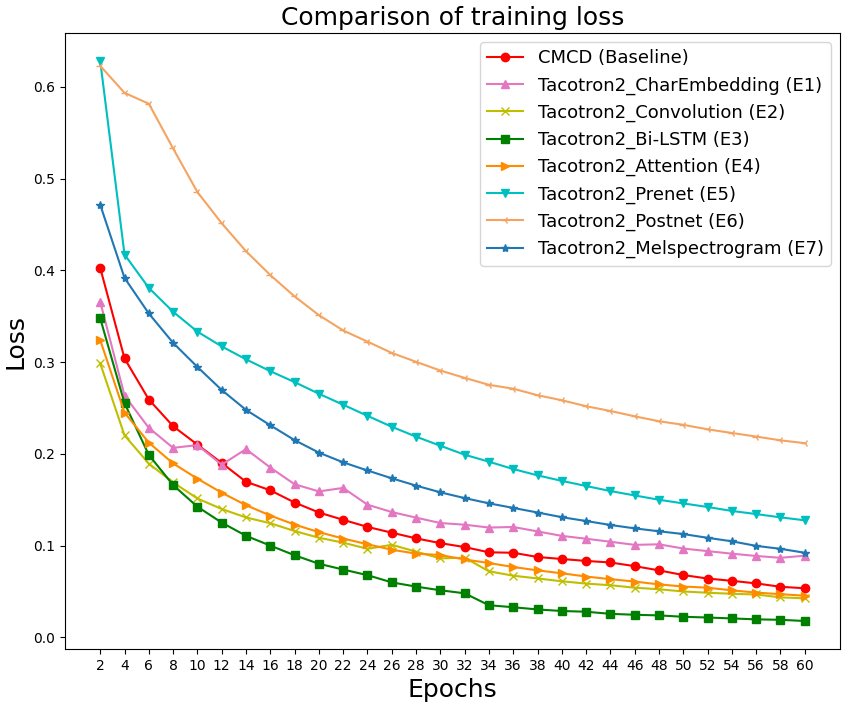}}
  \caption{Comparison of training loss - CMCD vs Tacotron 2 representations}
  \label{fig: training loss}
  \vspace{-0.5cm}
\end{figure}

Additionally, an interesting observation was made that the proposed network, utilizing intermediate representations extracted from E3, exhibits faster convergence, as illustrated in Fig.~\ref{fig: training loss}.

\section{Conclusion}
\label{sec:Conclusion}

This study presented an end-to-end architecture for open vocabulary keyword spotting that leverages insights from a pre-trained TTS system. The proposed approach exhibited competitive performance compared to baseline methods across four different datasets. Notably, it showcases its potential in distinguishing similar pronunciations of audio-text pairs in the Libriphrase hard dataset. The ablation study on intermediate layers of the Tacotron~2 model revealed that E3 (Bi-LSTM block output) exhibited the best performance and faster convergence during training. Moreover, the proposed approach showed consistent performance in keyword identification regardless of the word lengths considered and demonstrated its robustness in the OOV condition. 
In future work, the focus will be on optimizing the knowledge transfer effectively by exploring effective strategies to utilize all intermediate layers of the TTS model rather than relying on a single layer. 

\printbibliography

@article{lopez2021deep,
  title={Deep spoken keyword spotting: An overview},
  author={L{\'o}pez-Espejo, Iv{\'a}n and Tan, Zheng-Hua and Hansen, John HL and Jensen, Jesper},
  journal={IEEE Access},
  volume={10},
  pages={4169--4199},
  year={2021},
  publisher={IEEE}
}

@inproceedings{lugosch2018donut,
  title={DONUT: CTC-based Query-by-Example Keyword Spotting},
  author={Lugosch, L. and Myer, S. and Tomar, V. S.},
  booktitle={NeurIPS Workshop on Interpretability and Robustness in Audio, Speech, and Language},
  year={2018},
  address={Montreal, Canada},
  month={December}
}

@inproceedings{tang2018deep,
  title={Deep residual learning for small-footprint keyword spotting},
  author={Tang, Raphael and Lin, Jimmy},
  booktitle={Proc. ICASSP},
  pages={5484--5488},
  year={2018},
  organization={IEEE}
}

@inproceedings{sainath15b_interspeech,
  author={Tara N. Sainath and Carolina Parada},
  title={{Convolutional neural networks for small-footprint keyword spotting}},
  year=2015,
  booktitle={Proc. Interspeech},
  pages={1478--1482},
}

@inproceedings{huang2021query,
  title={Query-by-example keyword spotting system using multi-head attention and soft-triple loss},
  author={Huang, Jinmiao and Gharbieh, Waseem and Shim, Han Suk and Kim, Eugene},
  booktitle={Proc. ICASSP},
  pages={6858--6862},
  year={2021},
  organization={IEEE}
}

@article{liu2021neural,
  title={Neural keyword confidence estimation for open-vocabulary keyword spotting},
  author={Liu, Zuozhen and Li, Ta and Zhang, Pengyuan},
  journal={Electronics Letters},
  volume={58},
  number={3},
  pages={133--135},
  year={2021},
  publisher={IET}
}

@inproceedings{nishu23_interspeech,
  author={Kumari Nishu and Minsik Cho and Devang Naik},
  title={{Matching Latent Encoding for Audio-Text based Keyword Spotting}},
  year=2023,
  booktitle={Proc. INTERSPEECH},
  pages={1613--1617}
}

@inproceedings{lee23d_interspeech,
  author={Yong-Hyeok Lee and Namhyun Cho},
  title={{PhonMatchNet: Phoneme-Guided Zero-Shot Keyword Spotting for User-Defined Keywords}},
  year=2023,
  booktitle={Proc. INTERSPEECH},
  pages={3964--3968}
}

@inproceedings{shin22_interspeech,
  author={Hyeon-Kyeong Shin and Hyewon Han and Doyeon Kim and Soo-Whan Chung and Hong-Goo Kang},
  title={{Learning Audio-Text Agreement for Open-vocabulary Keyword Spotting}},
  year=2022,
  booktitle={Proc. INTERSPEECH},
  pages={1871--1875}
}

@inproceedings{sacchi2019open,
  title={Open-vocabulary keyword spotting with audio and text embeddings},
  author={Sacchi, Niccolo and Nanchen, Alexandre and Jaggi, Martin and Cernak, Milos},
  booktitle={INTERSPEECH 2019-IEEE International Conference on Acoustics, Speech, and Signal Processing},
  year={2019}
}

@inproceedings{shen2018natural,
  title={Natural tts synthesis by conditioning wavenet on mel spectrogram predictions},
  author={Shen, Jonathan and Pang, Ruoming and Weiss, Ron J and Schuster, Mike and Jaitly, Navdeep and Yang, Zongheng and Chen, Zhifeng and Zhang, Yu and Wang, Yuxuan and Skerrv-Ryan, Rj and others},
  booktitle={Proc. ICASSP},
  pages={4779--4783},
  year={2018},
  organization={IEEE}
}

@misc{NVIDIATacotron2,
  title = {Pretrained Tacotron2 model},
  author = {{NVIDIA Corporation}},
  howpublished = {\url{https://github.com/NVIDIA/tacotron2}},
}

@misc{ito2017ljspeech,
  author = {Keith Ito},
  title = {The LJ Speech Dataset},
  howpublished = {\url{https://keithito.com/LJ-Speech-Dataset/}},
  year = {2017}
}

@inproceedings{huang20i_interspeech,
  author={Wen-Chin Huang and Tomoki Hayashi and Yi-Chiao Wu and Hirokazu Kameoka and Tomoki Toda},
  title={{Voice Transformer Network: Sequence-to-Sequence Voice Conversion Using Transformer with Text-to-Speech Pretraining}},
  year=2020,
  booktitle={Proc. Interspeech},
  pages={4676--4680},
}

@article{vaswani2017attention,
  title={Attention is all you need},
  author={Vaswani, Ashish and Shazeer, Noam and Parmar, Niki and Uszkoreit, Jakob and Jones, Llion and Gomez, Aidan N and Kaiser, {\L}ukasz and Polosukhin, Illia},
  journal={Advances in neural information processing systems},
  volume={30},
  year={2017}
}

@inproceedings{panayotov2015librispeech,
  title={Librispeech: an asr corpus based on public domain audio books},
  author={Panayotov, Vassil and Chen, Guoguo and Povey, Daniel and Khudanpur, Sanjeev},
  booktitle={Proc. ICASSP},
  pages={5206--5210},
  year={2015},
  organization={IEEE}
}

@article{warden2018speech,
  title={Speech commands: A dataset for limited-vocabulary speech recognition},
  author={Warden, Pete},
  journal={arXiv preprint arXiv:1804.03209},
  year={2018}
}

@inproceedings{kim2019query,
  title={Query-by-example on-device keyword spotting},
  author={Kim, Byeonggeun and Lee, Mingu and Lee, Jinkyu and Kim, Yeonseok and Hwang, Kyuwoong},
  booktitle={Proc. ASRU},
  pages={532--538},
  year={2019},
  organization={IEEE}
}

@inproceedings{levenshtein1966binary,
  title={Binary codes capable of correcting deletions, insertions, and reversals},
  author={Levenshtein, Vladimir I and others},
  booktitle={Soviet physics doklady},
  volume={10},
  number={8},
  pages={707--710},
  year={1966},
  organization={Soviet Union}
}

@article{kingma2014adam,
  title={Adam: A method for stochastic optimization},
  author={Kingma, Diederik P and Ba, Jimmy},
  journal={arXiv preprint arXiv:1412.6980},
  year={2014}
}

@inproceedings{kgurugubelli2024comparative,
  title={Comparative Study of Tokenization Algorithms for End-to-End Open Vocabulary Keyword Detection},
  author={Gurugubelli, Krishna and Mohamed, Sahil and KS, Rajesh Krishna},
  booktitle={Proc. ICASSP},
  pages={12431--12435},
  year={2024},
  organization={IEEE}
}






\end{document}